\documentclass[twocolumn, secnumarabic, amsmath, amssymb, nobibnotes, aps, pra]{revtex4-2}

\usepackage{graphicx}
\usepackage{dcolumn}
\usepackage{bm}

\usepackage{upgreek}
\usepackage{amsmath}
\usepackage{times}
\usepackage{braket}
\usepackage{xcolor}
\usepackage{ulem}
\usepackage{soul}
\usepackage[colorlinks=True, linkcolor=blue, urlcolor=blue]{hyperref}

\begin{document}

\preprint{APS/123-QED}

\title{Optimizing brightness of SPDC source in Laguerre-Gaussian modes using type-0 periodically-poled nonlinear crystal}

\author{Jungmo Lee}
\author{Kyungdeuk Park}
\author{Dongkyu Kim}
\author{Yonggi Jo}
\author{Dong-Gil Im}
\email[Dong-Gil Im: ]{eastgil@add.re.kr}
\author{Yong Sup Ihn}
\email[Yong Sup Ihn: ]{yong0862@add.re.kr}
\affiliation{Ageny for Defense Development, Daejeon 34186, Korea} 
\date{\today}

\begin{abstract}
	Photon pairs generated via spontaneous parametric down-conversion (SPDC) can exhibit entanglement in the Laguerre-Gaussian (LG) mode basis, which enables high-dimensional free-space quantum communication by exploiting the high-dimensional space spanned by the LG modes. For such free-space quantum communication, the brightness of the quantum light source plays an important role due to the atmospheric turbulence and photon loss. A variety of studies have analyzed the SPDC brightness by decomposing biphoton states into LG modes, but they have often relied on a degenerate state, a narrow spectral bandwidth approximation, or a thin crystal approximation. However, these approaches are unsuitable for non-degenerate type-0 SPDC with a periodically-poled nonlinear crystal, which offers higher brightness due to its superior nonlinear coefficients. In this study, we examine the spectrum of photon pairs in specific LG modes generated by a type-0 ppKTP crystal while avoiding the constraints imposed by the aforementioned assumptions. In addition, we investigate the optimal focal parameters of the pump, signal, and idler to maximize the brightness for a given LG mode. Our findings show that it is not feasible to simultaneously optimize the brightness for different LG modes with a single pump focal parameter. The results of this study provide a comprehensive framework for developing high-brightness quantum light sources and contribute to the advancement of high-dimensional free-space quantum communication.
	
\end{abstract}
\maketitle

\section{\label{sec:level1}Introduction}

	Spontaneous parametric down-conversion (SPDC) is a pivotal process in quantum optics, where pump photons interact with a nonlinear crystal to generate two photons with lower frequencies, commonly referred to as signal and idler photons which can exhibit quantum correlations \cite{Klyshko88}. Due to their inherent quantum correlations and ease of generation, these entangled photon pairs have become the cornerstone of various quantum information processing applications, including quantum communication \cite{Gisin07,Ursin07}, quantum computing \cite{Kok07,Yao12}, and quantum metrology \cite{Giovannetti11}. In particular, SPDC serves as a reliable and efficient method for generating entangled photon pairs, enabling secure communication through quantum key distribution and playing a crucial role in the advancement of quantum communication technologies \cite{Ekert91,Yin20}.

	One of the key features of photon pairs generated via SPDC is their ability to exhibit entanglement in the Laguerre-Gaussian (LG) mode basis \cite{Miat11}. LG states provide an infinite-dimensional Hilbert space, allowing the encoding of information in high-dimensional quantum states \cite{Cozzo19Adv}. This capability enhances the channel capacity and offers resilience to noise compared to two-dimensional systems, which provides advantages for high-dimensional free-space quantum communication \cite{Seb19}. These benefits have been experimentally demonstrated in both laboratory environments and long-distance free-space channels, highlighting the versatility and potential of SPDC-based quantum light sources \cite{Vall14, Kren16, Pang18}.
		
	For free-space quantum communication, the brightness of the quantum light source plays an important role due to the atmospheric turbulence and photon loss. A high-brightness SPDC source increases the photon flux, improving reliable transmission over extended distances \cite{Fedr12, Lim23, Kim22}. Therefore, to effectively utilize the advantages of LG states in high-dimensional free-space quantum communications, it is essential to investigate the brightness of quantum light sources for specific LG modes, since different LG modes can exhibit varying efficiencies in the SPDC process \cite{Miat11}.
	
	A variety of studies have analyzed biphoton states by decomposing them into LG modes \cite{Pala11, Dixon14, Ben10, Ljung05, Ling08, Drag04, Torr03, Bag22, Sevi24, Yao11J, Sala11}, facilitating the optimization of the beam waist to maximize the brightness of quantum light sources \cite{Pala11, Dixon14, Ben10, Ljung05, Ling08, Drag04}. These studies typically rely on at least one of the following three assumptions: (i) a degenerate state, where the frequencies of signal and idler photons are identical \cite{Pala11, Torr03}, (ii) a narrow spectral bandwidth approximation, where the spectral distributions of signal and idler photons are sufficiently narrow \cite{Dixon14, Ben10, Ljung05, Bag22, Sevi24}, and (iii) a thin crystal approximation, where the focal parameter of the pump beam is sufficiently small relative to the nonlinear crystal length \cite{Ling08, Drag04, Yao11J, Sala11}. However, these assumptions are not suitable for non-degenerate type-0 SPDC, particularly in periodically-poled (pp) nonlinear crystals, where type-0 nonlinear crystal enables efficient use of the largest nonlinear coefficient, resulting in significant higher photon-pair brightness \cite{Stein14, Jab17, Park24}.
	
	In this work, we explore the spectral spectrum for specific LG modes of photon pairs using a type-0 ppKTP crystal, avoiding the common constraints of the aforementioned three conditions. In addition, we analyze the optimal pump, signal, and idler focal parameters to maximize the coincidence probability, providing a comprehensive framework for optimizing the brightness of SPDC photon pairs. Our investigation underscores the unique advantages of type-0 ppKTP crystals, which provide higher brightness due to their superior nonlinearity coefficients. We believe that our results will advance the development of high-brightness quantum light sources not only for high-dimensional free-space quantum communication but also for wavelength-multiplexed quantum communication utilizing the frequency-correlated photon pairs generated from the type-0 SPDC process \cite{Soren18,Kim21}.

\section{\label{sec:level2} Spectrum of photon pairs for a specific LG mode using a type-0 ppKTP}

	In this section, we derive the coincidence amplitude without imposing the aforementioned assumptions, and demonstrate the reduction in the number of dependent variables. The detailed derivation can be found in Appendix \ref{Appendix:B}.

	We start with a biphoton state of SPDC in the momentum space with a paraxial approximation. In the paraxial regime, the wave vector $\bm{k}$ can be decomposed into a transverse component $\bm{q}$ and a longitudinal component $k_z$, such that $\bm{k}=\bm{q}+k_z \bm{z}$. The wave number $k$ is defined as $k=|\bm{k}|$. Then, the biphoton state can be described as \cite{Saleh00},
			\begin{align}
				|\psi_{\text{SPDC}}\rangle =& \iint \bm{d q_s} \, \bm{d q_i} \, d\omega_{s}\,d\omega_{i}\, \, \Psi(\bm{q_s}, \bm{q_i}; \omega_{s}, \omega_{i}) \nonumber\\
				& \times \hat{a}_s^\dagger(\bm{q_s}, \omega_s) \, \hat{a}_i^\dagger(\bm{q_i}, \omega_i) \, |\text{vac}\rangle,
                \label{e2_1}
			\end{align}
    where $\omega$ is a frequency, $\hat{a}^\dagger$ is a creation operator, $|\text{vac}\rangle$ is a vacuum state, and $\Psi(\bm{q_s}, \bm{q_i}; \omega_{s}, \omega_{i})$ is a biphoton mode function. Here, we assume that the center of the nonlinear crystal is positioned at the origin of the coordinate system with its longitudinal direction aligned along the $z$ axis, and the signal and idler photons travel in directions nearly parallel to the pump propagation direction, known as the quasicollinear regime \cite{Bag21}. Considering the energy conservation ($\omega_p = \omega_s+\omega_i$) and the transverse wave vector conservation ($\bm{q_p}=\bm{q_s}+\bm{q_i}$) \cite{Saleh00}, the biphoton mode function is given as
			\begin{align}
                \Psi(\bm{q_s}, \bm{q_i}; \omega_s, \omega_i) =& \, \widetilde{V}_p(\bm{q_s+q_i}) S_p(\omega_s+\omega_i) \notag\\
                &\times \int_{-\frac{L}{2}}^{\frac{L}{2}} \,dz \,\exp(i\Delta k_z z),
                \label{e2_2}
			\end{align}
    where $\widetilde{V}_p(\bm{q_s+q_i})$ represents the spatial distribution of the pump beam in terms of angular spectrum, $S_p(\omega_s+\omega_i)$ represents the spectral distribution of the pump beam, and $L$ denotes the longitudinal length of the crystal. 
    The longitudinal wave vector mismatch $\Delta k_z$ is defined as $\Delta k_z=k_{pz}-k_{sz}-k_{iz}-\frac{2\pi}{\Lambda}$ for the periodically-poled (pp) crystal, where $\Lambda$ is the poling period.

	Now, to investigate the spectrum for a specific LG mode of photon pairs, we decompose the biphoton state into LG modes with eigenstates of LG basis, $|n_k,l_k,\omega_k\rangle$, where $n$ and $l$ denote the radial and azimuthal index, respectively \cite{Fick12}. The subscript $k$ stands for $p$ (pump), $s$ (signal), and $i$ (idler). Then, the biphoton state can be expressed as
			\begin{align}
                |\psi_{\text{SPDC}}\rangle =& \iint d\omega_{s}\,d\omega_{i}\, \sum_{n_s,n_i=0}^{\infty}\,\sum_{l_s,l_i=0}^{\infty} \notag \\  
                &\times C_{n_s,n_i}^{l_s, l_i}|n_s,l_s,\omega_s\rangle|n_i,l_i,\omega_i\rangle,
                \label{e5_4}
			\end{align}
    where $C_{n_s,n_i}^{l_s, l_i}$ is the coincidence amplitude. The eigenstate $|n_k,l_k,\omega_k\rangle$ is defined as \cite{Bag22}
	\begin{align}
                |n_k,l_k,\omega_k\rangle = \iint \bm{dq}\, \widetilde{LG}^{l_k}_{n_k}(\bm{q})\hat{a}^\dagger(\bm{q},\omega_k)|\text{vac}\rangle,
                \label{e5_5}
			\end{align}
    where the normalized amplitude of a specific LG mode in $\bm{k}$-space, denoted by $\widetilde{LG}^{l_k}_{n_k}$, is defined as follows \cite{Song20}
			\begin{align}
                \widetilde{LG}^{l_k}_{n_k}(\bm{q}) =& \,(-1)^{n_k}\sqrt{\frac{w^2n_k!}{8\pi^3(n_k+|l_k|)!}}\left(\frac{|\bm{q}|w_k}{\sqrt{2}}\right)^{|l_k|} \notag \\
		&\times L_{n_k}^{|l_k|}\left(\frac{|\bm{q}|^2w_k^2}{2}\right) \exp\left(-\frac{|\bm{q}|^2w_k^2}{4}\right) \notag \\
                &\times \text{exp}\left(il_k\left(\varphi-\frac{\pi}{2}\right)\right),
                \label{e5_1}
			\end{align}
    where $L_{n_k}^{|l_k|}$ represents the associated Laguerre polynomial, $w$ is the beam waist, and $\varphi$ is the azimuthal angle in $\bm{k}$-space. Then, the coincidence amplitude can be obtained by projecting the biphoton state onto an eigenstate corresponding to a specific LG mode,   
            \begin{align}
                C_{n_s,n_i}^{l_s, l_i} =\,& \langle n_s,l_s,\omega_s ; n_i,l_i,\omega_i|\psi_{\text{SPDC}}\rangle \notag \\ 
                =&\iint \bm{dq_s}\bm{dq_i}\, S_p(\omega_s+\omega_i)\widetilde{V_p}(\bm{q_s}+\bm{q_i}) \notag \\  
                &\times [\widetilde{LG}^{l_s}_{n_s} (\bm{q_s})]^*[\widetilde{LG}^{l_i}_{n_i} (\bm{q_i})]^*\int^{\frac{L}{2}}_{-\frac{L}{2}} dz\,\exp(i\Delta k_z z).
                \label{e5_6}
            \end{align}
    Equation~(\ref{e5_6}) can be further simplified under three specific conditions: (i) the degenerate state, in which the signal and idler photons possess identical frequencies; (ii) the narrow spectral bandwidth approximation, in which the spectral distributions of the signal and idler photons are sufficiently narrow; and (iii) the thin-crystal approximation, in which the focal parameter of the pump beam is sufficiently small.
    Here, the focal parameter is defined as the ratio of the nonlinear crystal length to the square of the pump beam waist \cite{Ben10}. 
    However, such simplifications are not applicable to the non-degenerate type-0 SPDC process in a periodically-poled (pp) nonlinear crystal (See Appendix \ref{Appendix:A} for details).
	
    To explicitly evaluate Eq.~(\ref{e5_6}) without relying on aforementioned approximations, we reformulate the coincidence amplitude using the inverse Fourier transform,
			\begin{align}
                C_{n_s,n_i}^{l_s, l_i} &= \frac{1}{4\pi^2}\int_0^{\infty} r\,dr \int_0^{2\pi} d\phi \, \int^{\frac{L}{2}}_{-\frac{L}{2}} dz \, S_p(\omega_s+\omega_i) \notag \\
                &\times V^{'}_p(r, z)[{LG^{'}}^{l_s}_{n_s}(r, z)]^*[{LG^{'}}^{l_i}_{n_i}(r, z)]^* \notag \\
                &\times \exp(i\Delta k \, z)\exp(i(l_p-l_s-l_i)\phi),
                \label{e5_7}
            \end{align}
    where $V^{'}_p(r,z)$ denotes the spatial distribution of the pump beam in the $\bm{x}$-space, and $\phi$ is the azimuthal angle in $\bm{x}$-space. 
    Under the definition of the angular spectrum and the paraxial approximation, the longitudinal wave vector mismatch $\Delta k_z$ in the exponential term is replaced by $\Delta k (=k_p-k_s-k_i-\frac{2\pi}{\Lambda})$. This substitution makes the phase mismatch independent of $q_s$ and $q_i$, which implies that it no longer depends on the propagation direction of the beams. As a result, this reformulation significantly simplifies the calculation of the coincidence amplitude. Furthermore, ${LG^{'}}^{l_k}_{n_k}(r,z)$ represents the radial ($r$) dependence of the normalized amplitude for a specific LG mode in $\bm{x}$-space and is formally defined in \cite{Vall17}
            \begin{align}
                {LG^{'}}^{l_k}_{n_k}(r, z) &=\, \sqrt{\frac{n_k!}{\pi(n_k+|l_k|)!}}\left(\frac{g_k^*}{g_k}\right)^n\exp\left(\frac{-r^2}{g_k w_k^2}\right) \notag \\ 
               \times & \left(\frac{\sqrt{2}}{g_k w_k}\right)^{|l_k|+1}r^{|l_k|}\,L_{n_k}^{|l_k|}\left(\frac{2r^2}{w_k^2|g_k|^2}\right), \notag \\ 
                \label{e5_2}
            \end{align}
    where the complex beam parameter $g_k$ is given by
            \begin{align}
                g_k = 1+i\frac{2z}{k_k w_k^2}=1+if_k u.
                \label{e5_3}
            \end{align}
    Here, the focal parameter $f_k = \frac{L}{k_kw_k^2}$ and $u=\frac{2z}{L}$. To simplify the calculation, we assume a monochromatic pump beam, in which the pump frequency $\omega_p$ is fixed, and the pump beam is in a Gaussian spatial mode. After integrating over the radial coordinate $r$ and azimuthal angle $\phi$, and applying the orbital angular momentum (OAM) conservation law $l_p = l_s + l_i=0$ \cite{Bag22}, the coincidence amplitude simplifies to
			\begin{align}
                C^l_{n_s, n_i}&=\sum_{m_s=0}^{n_s}\sum_{m_i=0}^{n_i}\alpha^{l, n_s, n_i}_{m_s, m_i} \int_{-1}^1\,du\,G_{m_s, m_i}^{l, n_s, n_i}(u)\,\exp(i\Phi u),
                \label{e5_9}
			\end{align}
    where $l = l_s = -l_i$, and $\alpha^{l, n_s, n_i}_{m_s, m_i}$ is a coefficient independent of $u$ but dependent on the beam waist $w_p$, $w_s$ and $w_i$. The phase mismatch parameter is given by $\Phi=\frac{\Delta k L}{2}$, and $G_{m_s, m_i}^{l, n_s, n_i}$ is given by
		\begin{align}
                G_{m_s, m_i}^{l, n_s, n_i} &= \frac{g_p^{m_s+m_i+l}g_s^{n_s-m_s}g_i^{n_i-m_i}}{{g_s^*}^{n_s-m_i}{g_i^*}^{n_i-m_s}{g^{*}(g_p, g_s, g_i)}^{m_s+m_i+l+1}},
                \label{e5_10}
		\end{align}
    with $g^{*}(g_p, g_s, g_i) = \frac{w_p^2w_s^2g_pg_s^* + w_p^2w_i^2g_pg_i^* + w_s^2w_i^2g_s^*g_i^*}{w_p^2w_s^2+w_p^2w_i^2+w_s^2w_i^2}$.

	Equation (\ref{e5_9}) enables direct calculation of $C^l_{n_s, n_i}$ without additional assumptions. However, when the temperature $T$ is fixed, the dependence on five variables ($f_p$, $f_s$, $f_i$, $w_s$, $w_i$) introduces a large parameter space, making optimization computationally demanding. To optimize brightness more efficiently, we introduce two conditions to reduce the number of dependent parameters in $C^l_{n_s, n_i}$. First, following the condition for maximizing the coincidence probability $P^l_{n_{si}} = |C^l_{n_{si}}|^2$ in a previous study \cite{Miat11}, we set the beam waists and radial indices for the signal and idler photons to be identical, such that $w_s=w_i=w_{si}$ and $n_s=n_i=n_{si}$. This approach maximizes the photon pair collection rate $R_c$. Second, the signal and idler wave number $k_s$ and $k_i$ deviate symmetrically from the degenerate wave number $k_d = k(\frac{\omega_p}{2})$, which corresponds to the wave number at degeneracy. This approach is valid for type-0 SPDC, where the signal and idler photons share the same polarization \cite{Stein14}, minimizing refractive index differences. Under these two conditions, $G_{m_s, m_i}^{l, n_s, n_i}$ can be approximated by ${G^d}^{l, n_{si}}_{m_s, m_i}$, where all parameters of $g_s$, $g_i$, and $g^{*}(g_p, g_s, g_i)$ are replaced by the degenerate complex beam parameter $g^{d}_{si} = 1 + i f^{d}_{si} u$, where the degenerate focal parameter $f^d_{si}$ is defined as $f^d_{si}=\frac{L}{k_d w_{si}^2}$. Consequently, the coincidence amplitude simplifies to
			\begin{align}
                C^l_{n_{si}} \approx \sum_{m_s=0}^{n_{si}}\sum_{m_i=0}^{n_{si}} \beta^{l, n_{si}}_{m_s, m_i} \int_{-1}^1\,du\,{G^d}^{l, n_{si}}_{m_s, m_i}(u)\,\exp(i\Phi u),
                \label{e5_14}
			\end{align}
    where ${G^d}^{l, n_{si}}_{m_s, m_i} = g_p^{m_s+m_i+l}{g^{d}_{si}}^{2n_{si}-m_s-m_i} / {{g^{d}_{si}}^*}^{2n_{si}+l+1}$, and $\beta^{l, n_{si}}_{m_s, m_i}$ is a coefficient independent of $u$ but dependent on $f_p$ and $f^d_{si}$. As a result, $C^l_{n_{si}}$ is determined solely by the pump focal parameter $f_p$ and the degenerate focal parameter $f^d_{si}$ when the temperature $T$ is fixed (see Appendix \ref{Appendix:B} for more details). 
    Note that we define this procedure as the degenerate approximation. Unlike the conventional degenerate phase matching condition adopted in previous studies which applies degeneracy to both the complex beam parameters and the phase mismatch parameter $\Phi$, our approximation restricts the degeneracy solely to the condition where the signal and idler modes share an identical beam waist and radial index, and their frequencies are symmetrically distributed with respect to half the pump frequency. Under this constraint, $g_s$, $g_i$, and $g^{*}(g_p, g_s, g_i)$ can be consistently replaced with $g^{d}_{si}$. Notably, in this approximation, degeneracy is not imposed on $\Phi$, distinguishing our approach from prior studies.
    
\begin{figure}[t!]
                \centering
                \includegraphics[width=0.45\textwidth]{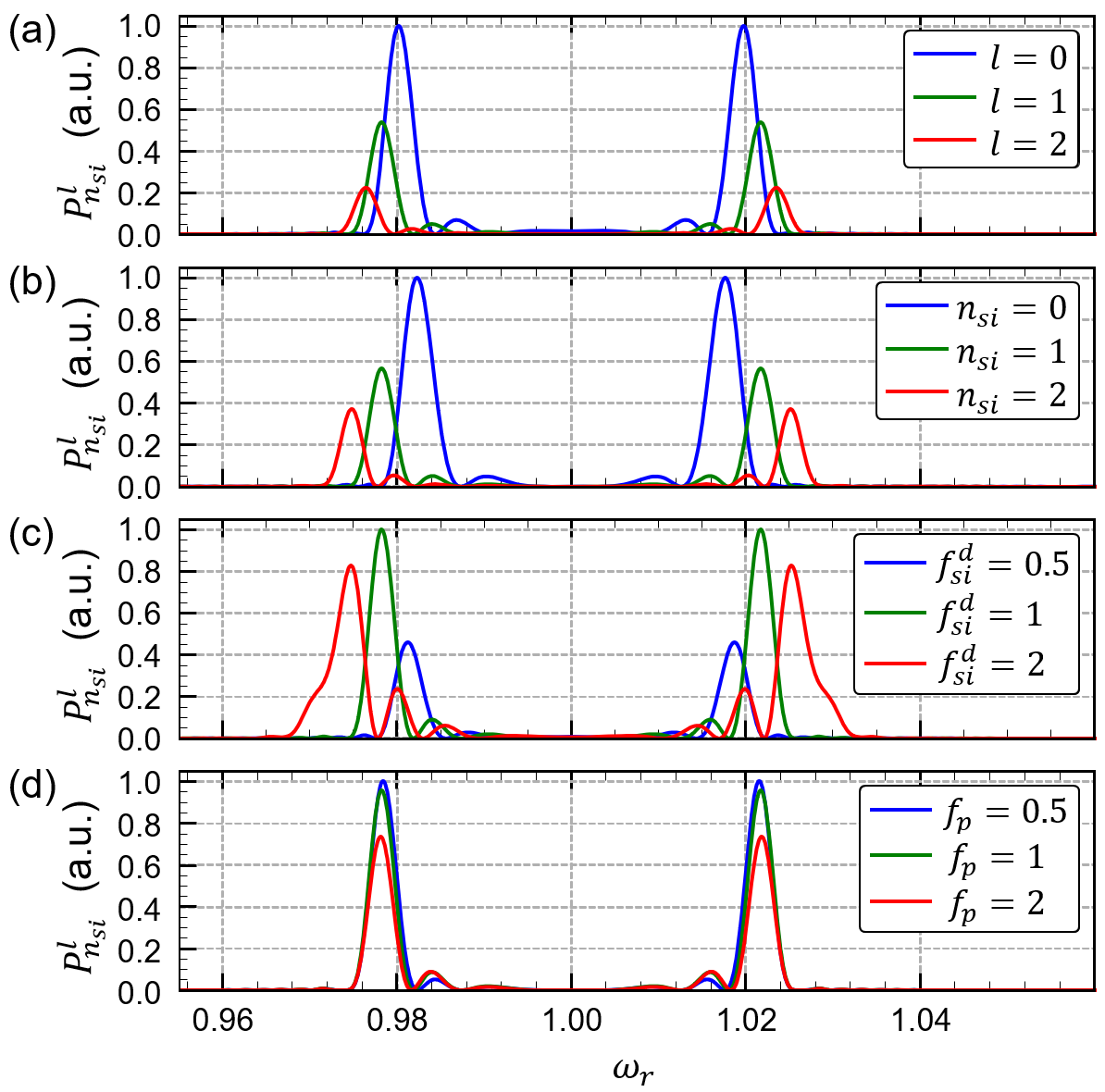}
                \caption{Coincidence probability $P^l_{n_{si}}$ as a function of the normalized frequency $\omega_r$, varying: (a) $l$ with $f_p = f^d_{si} = 1$, $n_{si} = 1$; (b) $n_{si}$ with $f_p = f^d_{si} = 1$, $l = 1$; (c) $f^d_{si}$ with $f_p = 1$, $l = n_{si} = 1$; and (d) $f_p$ with $f^d_{si} = 1$, $l = n_{si} = 1$. 
			     Each plot is normalized so that the global maximum of $P^l_{n_{si}}$ is set to unity. Here, the temperature $T$ is set to 24.5$\,^{\circ}\mathrm{C}$.}
                \label{fig01}
            \end{figure}

    To evaluate Eq.~(\ref{e5_14}), we consider a type-0 periodically-poled potassium titanyl phosphate (ppKTP) crystal with a length $L$ = 30 $\mathrm{mm}$ and a poling period $\Lambda$ = 3.425 $\mathrm{\upmu m}$ at $T = 24.5 \,^{\circ}\mathrm{C}$. The crystal is pumped by a continuous-wave laser at 405 nm. Figure \ref{fig01} presents the coincidence probability $P^l_{n_{si}}$ as a function of the normalized frequency of the down-converted photons, where $\omega_r=\frac{2\omega}{\omega_p}$. Each subplot in Fig. \ref{fig01} illustrates the effect of varying a single parameter ($l$, $n_{si}$, $f^d_{si}$, or $f_p$). As shown in Fig.~\ref{fig01}(a), \ref{fig01}(b), and \ref{fig01}(c), increasing $l$, $n_{si}$, or $f^d_{si}$ causes the spectral distribution of $P^l_{n_{si}}$ to shift away from the central frequency $\omega_r$ = 1.
    This shift occurs because an increase in LG indices and focal parameters reduces $k_z$, thereby compensating for the phase mismatch at wavelengths farther from the central frequency \cite{Hua18, Sevi24}. In contrast, Figure \ref{fig01}(d) shows that variations in $f_p$ have a relatively minor impact on the spectral distribution of $P^l_{n_{si}}$ compared to variations in $f^{d}_{si}$. This result is consistent with the scenario in which the pump, signal, and idler modes are all Gaussian \cite{Varga22}.

\section{\label{sec:level3}Pair collection rate and optimization of focal parameters}

	To discuss the brightness of SPDC photon pairs for a specific LG mode, we define the pair collection rate, $R_c$, as the integral of the coincidence probability $P^{l}_{n_{si}}$ over the signal-idler frequency $\omega_{si}$, given by
		\begin{align}
                R_c = \int d\omega_{si}\,P^l_{n_{si}}.
       \label{pcr}
		\end{align}
    Pair collection rate $R_c$ not only quantifies the coincidence probability for a given LG mode, but it can also be experimentally obtained by measuring the coincidence count rates using a spatial light modulator (SLM) and a single-mode fiber \cite{Arlt98}. Since $R_c$ depends on the pump focal parameter $f_p$ and the degenerate focal parameter $f^d_{si}$ at a fixed temperature, we examine $R_c$ as a function of these parameters for various LG indices (see Appendix \ref{Appendix:B} for more details). 
    
\begin{figure}[t!]
                \centering
                \includegraphics[width=0.45\textwidth]{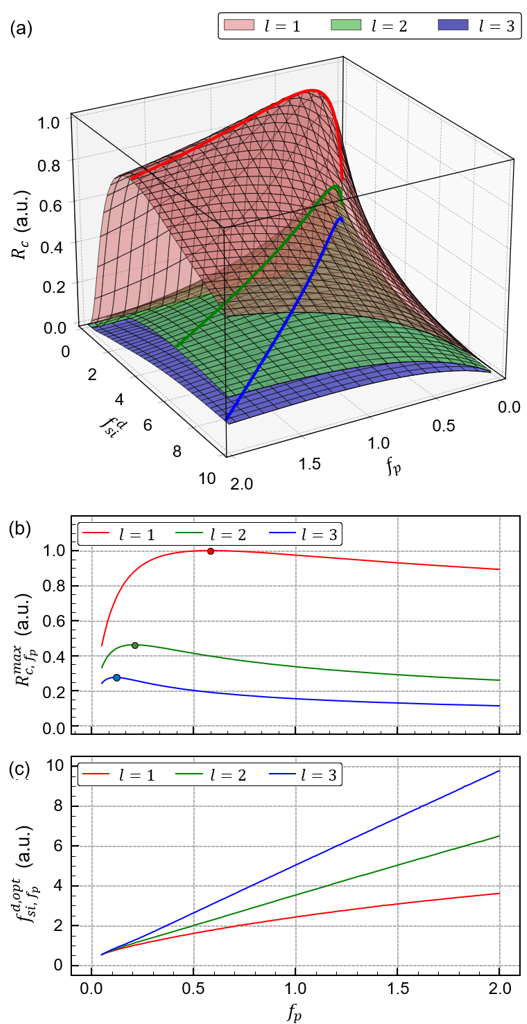}
                \caption{(a) Pair collection rate $R_c$ as a function of the degenerate focal parameter $f^d_{si}$ and the pump focal parameter $f_p$ for three values of $l = 1, 2 $ and $3$ with $n_{si} = 0$ at $T = 24.5\,^{\circ}\mathrm{C}$. The solid lines indicate $R_{c,f_p}^{max}$ and $f_{si,f_p}^{d, opt}$. (b) $R_{c,f_p}^{max}$ as a function of the pump focal parameter $f_p$. The solid dots represent the maximized pair collection rate $R_c^{max}$ and the optimized pump focal parameter $f_p^{opt}$. (c) $f_{si, f_p}^{d,opt}$ as a function of the pump focal parameter $f_p$.}
                \label{fig02}
            \end{figure}

    We first investigate the dependence of the pair collection rate on the degenerate focal parameter $f_{si}^{d}$ and the pump focal parameter $f_p$ for varying azimuthal index $l$ with $n_{si} = 0$, as shown in Fig.~\ref{fig02}(a). Since the pair collection rate exhibits a maximum for a fixed $f_p$, we define the maximum pair collection rate for a given $f_p$ as $R_{c,f_p}^{max}=\underset{f_{si}^{d}}{\max}\, R_c(f_p, f_{si}^{d})$, and the degenerate focal parameter that maximizes the pair collection rate for a given $f_p$ as $f_{si,f_p}^{d, opt}=\underset{f_{si}^{d}}{\arg\max}\, R_c(f_p, f_{si}^{d})$. The solid lines in Fig.~\ref{fig02}(a) indicate $R_{c,f_p}^{max}$ and $f_{si,f_p}^{d, opt}$. To further clarify this dependence, we plot $R_{c,f_p}^{max}$ and $f_{si,f_p}^{d, opt}$ as functions of $f_p$ in Fig.~\ref{fig02}(b) and \ref{fig02}(c), respectively. As shown in Fig.~\ref{fig02}(b), $R_{c,f_p}^{max}$ reaches a maximum, indicated by solid dots, at an optimized pump focal parameter $f_p^{opt} = \underset{f_p}{\arg\max}\, R_{c,f_p}^{max}$ with the corresponding maximized pair collection rate $R_{c}^{max} = \underset{f_p}{\max}\, R_{c,f_p}^{max}$.
    Especially, as $l$ increases, both $f_p^{opt}$ and $R_{c}^{max}$ decrease. Figure~\ref{fig02}(c) shows that as $l$ increases, $f_{si,f_p}^{d, opt}$ increases for the same $f_p$.

            \begin{figure}[t!]
                \centering
                \includegraphics[width=0.45\textwidth]{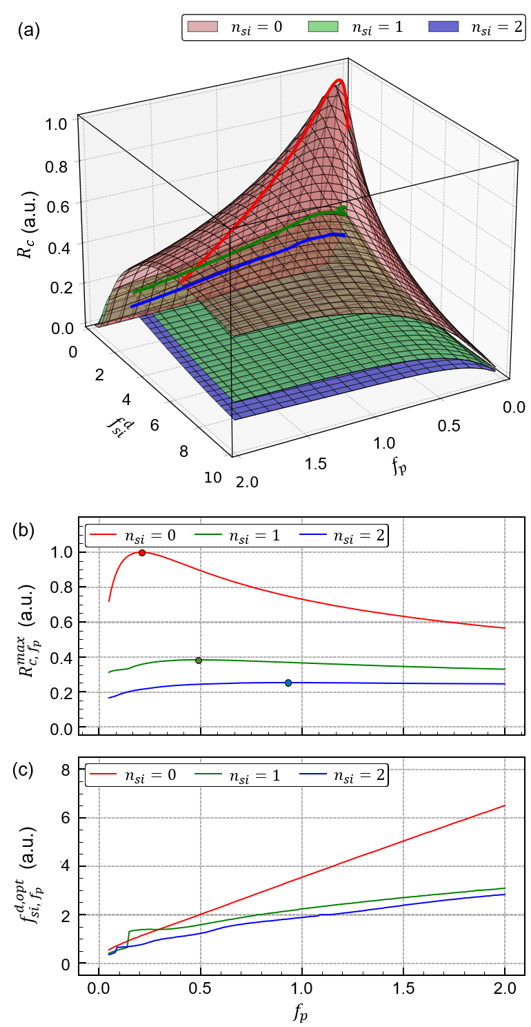}
                \caption{(a) Pair collection rate $R_c$ as functions of the degenerate focal parameter $f^d_{si}$ and the pump focal parameter $f_p$ for three values of $n_{si} = 0, 1 $ and $2$ with $l = 2$ at $T = 24.5 \,^{\circ}\mathrm{C}$. The solid lines indicate $R_{c,f_p}^{max}$ and $f_{si,f_p}^{d, opt}$. (b) $R_{c,f_p}^{max}$ as a function of the pump focal parameter $f_p$. The solid dots represent the maximized pair collection rate $R_c^{max}$ and the optimized pump focal parameter $f_p^{opt}$. (c) $f_{si, f_p}^{d,opt}$ as a function of the pump focal parameter $f_p$.}
                \label{fig03}
            \end{figure}

	Next, we examine the pair collection rate by varying the radial index $n_{si}$ for a fixed azimuthal index $l=2$, as shown in Fig.~\ref{fig03}(a). The solid lines in Fig.~\ref{fig03}(a) indicate $R_{c,f_p}^{max}$ and $f_{si,f_p}^{d, opt}$. As shown in Fig.~\ref{fig03}(b), increasing $n_{si}$ results in an increase in $f_p^{opt}$, while the maximized pair collection rate $R_{c}^{max}$ decreases. Figure~\ref{fig03}(c) shows that as $n_{si}$ increases, $f_{si,f_p}^{d, opt}$ for the same $f_p$ decreases. These results indicate that $R_c$ decreases with increasing $l$ and $n_{si}$, suggesting that higher-order LG modes exhibit larger divergence angles, leading to greater phase mismatching and ultimately reduced brightness \cite{Hua18}. The effect of $l$ and $n_{si}$ on {$f_{si, f_p}^{d, opt}$} arises from their impact on the photon distribution in $k$-space. As $l$ increases, the photons become more shift away from the center of $k$-space \cite{Pad15}. Since photons far from the center of $k$-space have frequencies further from the central frequency \cite{Yin21}, $f_{si}^{d}$ should be increased to efficiently collect those photons (see Fig.~\ref{fig01}(c)). Consequently, as $l$ increases, $f_{si, f_p}^{d, opt}$ should also increase to achieve $R_c^{max}$. In the same manner, as $n_{si}$ increases, the spatial distribution of photons becomes more concentrated near the center of $k$-space \cite{Miat11}, leading to a decrease in $f_{si, f_p}^{d, opt}$ required to achieve $R_c^{max}$. Since $f_{si, f_p}^{d, opt}$ is influenced not only by the LG indices but also by temperature $T$, we further investigate the effect of $T$ on the pair collection rate. Detailed results on $T$ dependence are provided in Appendix \ref{Appendix:C}.

            \begin{figure}[!t]
                \centering
                \includegraphics[width=0.43\textwidth]{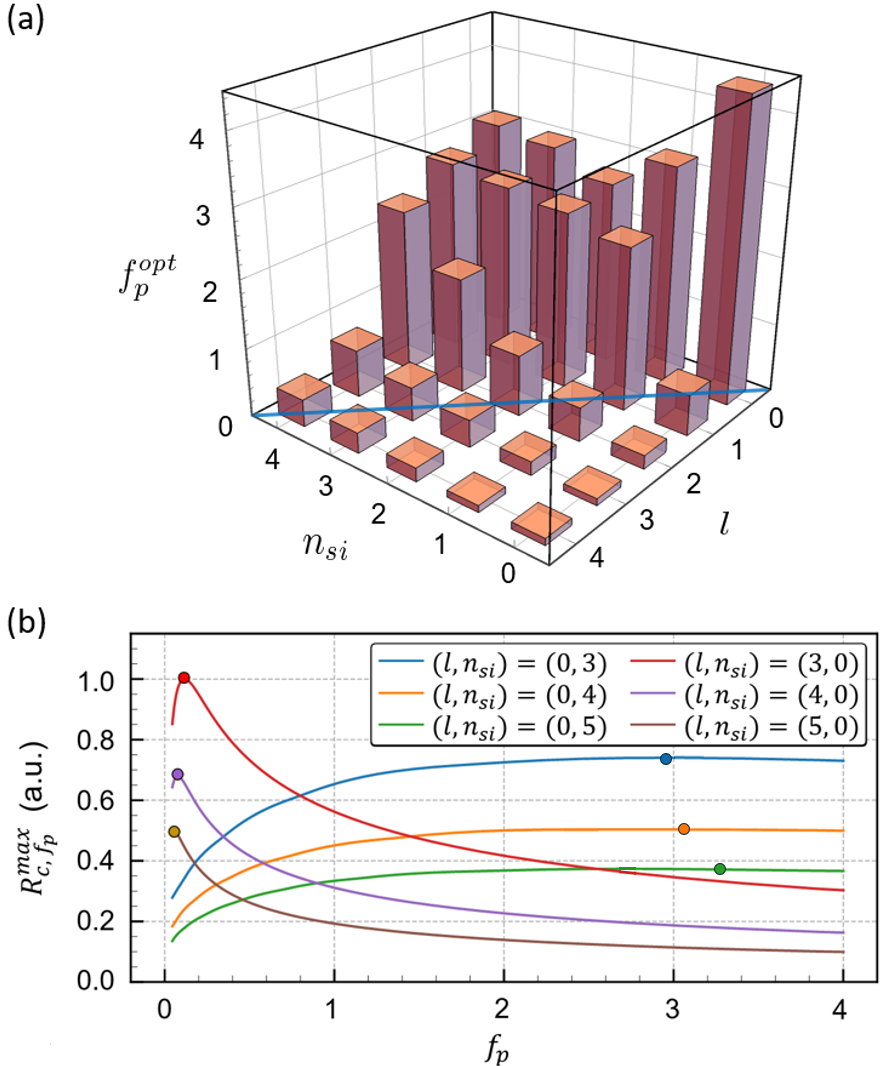}
                \caption{(a) Optimized pump focal parameter $f_p^{opt}$ as a function of the different Laguerre-Gaussian (LG) modes with $l,n \leq 4$ at $T = 24.5 \,^{\circ}\mathrm{C}$. The blue solid line indicates the condition $l=n_{si}$. (b) $R_{c,f_p}^{max}$ as a function of the pump focal parameter $f_p$ for different LG mode indices: $(l,n_{si}) = (0, 3), (0, 4), (0, 5), (3, 0), (4, 0)$ and $(5, 0)$ at $T = 24.5 \,^{\circ}\mathrm{C}$. The solid dots indicate the maximized pair collection rates $R_c^{max}$ and the optimized pump focal parameter $f_p^{opt}$.}
                \label{fig04}
            \end{figure}

	Based on our investigations, we discuss the experimental implementation of various LG modes of down-converted photons for high-dimensional quantum communication. Photon pairs generated via the SPDC process can be spatially separated for different LG modes using a spatial light modulator (SLM) \cite{Arlt98}. Each spatially separated photon possesses distinct focal parameters, which can be collected using an appropriate set of lenses tailored to their respective focal parameters \cite{Sevi24}. Meanwhile, the pump focal parameter $f_{p}$ is predetermined before the SPDC process, making independent control impossible for individual LG modes of photons. Therefore, in experimental implementations, $f_p$ is fixed at a specific value for the entire set of LG modes, while the degenerate focal parameter $f_{si}^d$ can be individually optimized for each LG mode. This limitation implies that it is not feasible to simultaneously optimize the pair collection rates for different LG modes using a single $f_p$. The discrepancy in $f_p^{opt}$ across different LG modes becomes more evident when comparing the cases of $l < n_{si}$ and $l > n_{si}$. 
    Figure~\ref{fig04}(a) shows the optimal focal parameter $f_p^{\mathrm{opt}}$ as a function of various LG modes with $l,n \leq 4$. The values of $f_p^{\mathrm{opt}}$ exhibit abrupt changes along the diagonal line defined by $l = n_{si}$, marked in solid blue line. Moreover, the difference in $f_p^{\mathrm{opt}}$ between the two regions increases as the index difference $|l - n_{si}|$ becomes larger. In particular, as shown in Fig.~\ref{fig04}(b), when $l = 0$ and $n_{si} = 3, 4, 5$ (i.e., $l - n_{si} \le -3$), $f_p^{\mathrm{opt}}$ remains below $0.2$. In contrast, for $n_{si} = 0$ and $l = 3, 4, 5$ (i.e., $l - n_{si} \ge 3$), $f_p^{\mathrm{opt}}$ exceeds $2$. Consequently, selecting $f_p^{opt}$ for $l-n_{si} \le -3$ to maximize the pair collection rate reduces the pair collection rate for $l-n_{si} \ge 3$ by more than half compared to its $R_c^{max}$.
    Therefore, when utilizing multiple LG modes simultaneously in an experiment, it is necessary to select appropriate LG modes based on the $f_p^{opt}$ of each mode. Furthermore, it is crucial to properly determine $f_p^{opt}$, taking into account the reduction in $R_c$ due to mismatches in $f_p^{opt}$ across different LG modes.

\section{\label{sec:level8}Conclusion}

In this study, we have systematically analyzed the spectral properties of specific LG modes in photon pairs generated via type-0 SPDC in a ppKTP crystal, without relying on the commonly used approximations, including the assumptions of the degenerate state, the narrow spectral bandwidth, and the thin crystal.
In addition, we have investigated the optimal conditions for the pump, signal, and idler focal parameters to maximize the pair collection rate for a specific LG mode. 
Our findings reveal the impossibility of simultaneously optimizing the pair collection rate across different LG modes using a single pump focal parameter.
The results presented in this work provide a comprehensive framework for designing high-brightness quantum light sources and offer critical insights for optimizing the pump focal parameter in practical implementations. 

Our study is not limited to maximizing the pair collection rate without relying on the three approximations previously discussed; it can also be extended to optimize the Schmidt number, which is a critical parameter in high-dimensional quantum communication \cite{Law04Sch}. 
In future work, we aim to develop an efficient method for calculating the Schmidt number in a high-dimensional space composed of a broad range of LG indices. This will enable systematic optimization of the Schmidt number through proper control of focal parameters. Moreover, recent studies have explored the enhancement of photon-pair purity and Schmidt number by employing arbitrary pump modes instead of Gaussian beams, as well as by engineering the nonlinear crystal domain \cite{Bag23Sch, Liu18Sch, Ber24Sch}. We envision to extend our approach by incorporating these techniques to further improve the Schmidt number. We believe that our results will contribute to the advancement of high-dimensional free-space quantum communication and wavelength-multiplexed quantum networks, leveraging frequency-correlated photon pairs generated through the type-0 SPDC process. 

\begin{acknowledgments}
     This work was supported by the Agency for Defense Development Grant funded by the Korean Government.
\end{acknowledgments}

\appendix
\onecolumngrid

\section{\label{Appendix:A} Limitations of previous approximation conditions}
\renewcommand{\thefigure}{A\arabic{figure}}
\setcounter{figure}{0}
	Previous studies analyzing biphoton states have commonly relied on one or more of the following approximation conditions: the degenerate state assumption, the narrow spectral bandwidth approximation, and the thin crystal approximation. 
However, these assumptions are not generally applicable to collinear type-0 pp crystals.
In this appendix, we outline the limitations of these approximations and highlight the necessity of a more general analytic framework.

\subsection{\label{Appendix:A-1} Degenerate state assumption}

	The degenerate state assumption, in which the signal and idler photons have identical frequencies (i.e., $\omega_s=\omega_i=\frac{\omega_p}{2}$), is not suitable for collinear type-0 pp crystals. 
In such systems, the signal and idler photons propagate along the same optical path and share the same polarization state.
As a result, conventional polarization beam splitters cannot be used for photon separation; instead, dichroic mirrors are employed to distinguish the signal and idler photons \cite{Stein14}. 
This constraint necessitates the consideration of a non-degenerate biphoton state, where the signal and idler photons have distinct frequencies.

            \begin{figure}[htbp]
            \centering
                \includegraphics[width=0.7\textwidth]{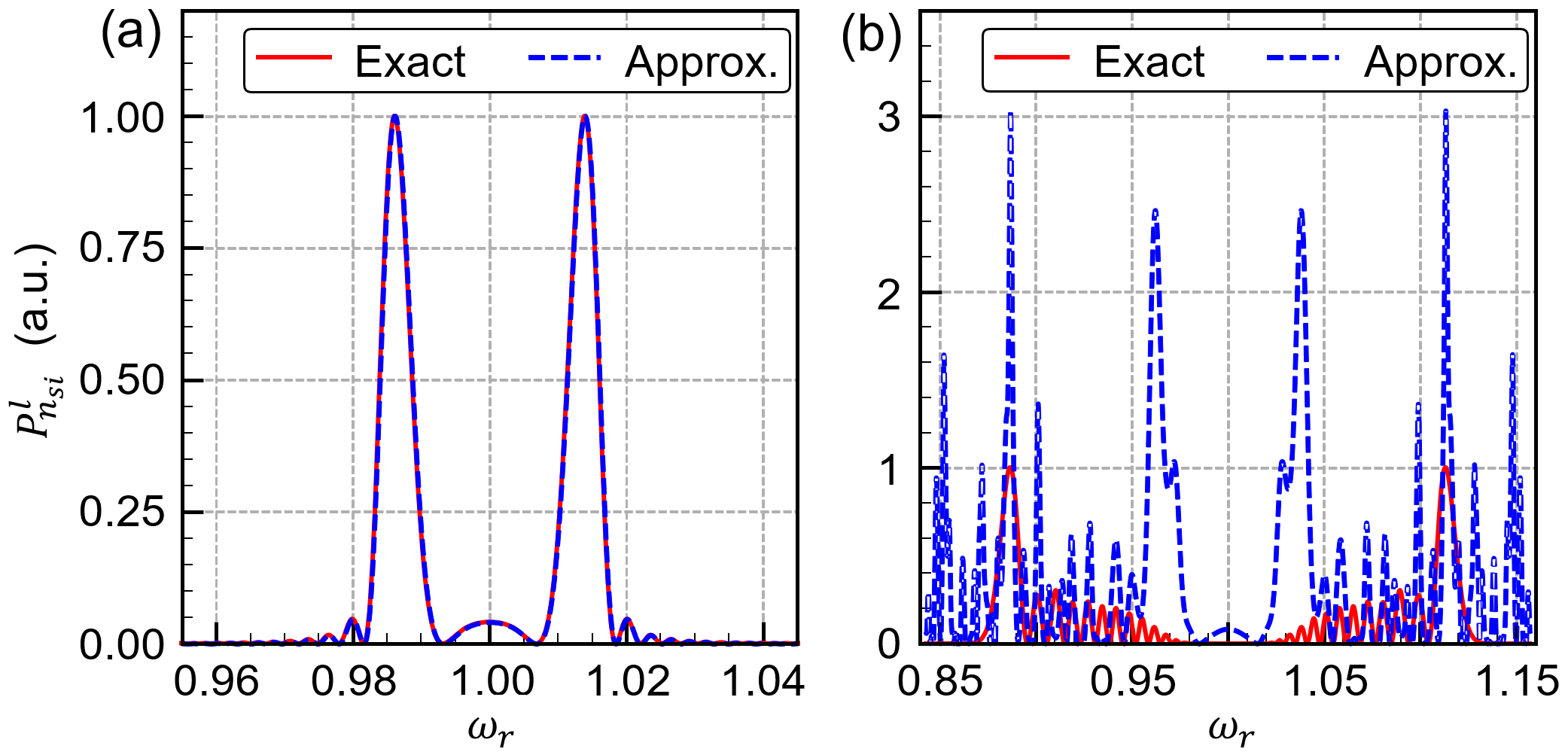}
                \caption{Normalized coincidence probability $P^{l}_{n_{si}}$ as a function of the normalized frequency $\omega_r$, comparing exact values (red solid line) and quadratic approximation (blue dashed line) under different conditions: (a) ($l,n_{si}$) = (0, 0), $(f_p,f^d_{si})$ = (0.1, 0.1). (b) ($l,n_{si}$) = (5, 5), $(f_p,f^d_{si})$ = (2, 10) at $T = 24.5\,^{\circ}\mathrm{C}$. The plots are normalized so that the global maxima of $P^{l}_{n_{si}}$ is set to unity.}
                \label{figa1}
            \end{figure}

\subsection{\label{Appendix:A-2} Narrow spectral bandwidth approximation}

	The narrow spectral bandwidth approximation assumes that the spectral distributions of signal and idler photons are sufficiently narrow, mathematically expressed as $\frac{\Delta \omega^{b}_{s\,(i)}}{\omega^{c}_{si}}\ll1$, where $\Delta \omega^{b}_{s\,(i)}$ represents the spectral bandwidth for the signal (or idler) photons, and $\omega^{c}_{s{i}}$ denotes the central frequency of the signal (or idler) photons. 
This condition can be extended to the wave number $k$ as $\frac{\Delta k^{b}_{s\,(i)}}{k^{c}_{si}}\ll1$, where $\Delta k^{b}_{s\,(i)}$ is the wave number bandwidth, and $k^{c}_{s{i}}$ is the central wave number of signal (or idler) photons. 
This approximation is valid when a spectral filter is employed \cite{Ljung05}; however, it inevitably reduces brightness, making it unsuitable for studies aiming to maximize brightness. 
Recent studies have addressed this issue by employing a quadratic approximation around the frequency satisfying $\Delta k = 0$ for $k_z$, eliminating the need for spectral filtering \cite{Bag22}.
While this approach has been successfully applied to type-II SPDC systems, where the biphoton spectrum is inherently narrow, its applicability to type-0 SPDC systems is limited.
Specifically, the quadratic approximation remains valid when the signal and idler focal parameters or the LG indices are small, as shown in Fig.~\ref{figa1}(a). 
However, as the focal parameters and LG indices increase, the spectral distribution of the coincidence probability shifts further from the center frequency (see Fig. \ref{fig01}(a), \ref{fig01}(b), \ref{fig01}(c) in the main text). 
Consequently, the deviation from the $\Delta k = 0$ condition increases, leading to greater errors in the quadratic approximation for $k_z$. 
As shown in Fig. \ref{figa1}(b), the discrepancy between the exact solution (red solid line) and the quadratic approximation (blue dashed line) increases as the focal parameters and LG indices grow. 
Therefore, to accurately analyze the coincidence probability for general focal parameters and LG indices, a method that does not rely on the narrow spectral bandwidth approximation is required.

\subsection{\label{Appendix:A-3} Thin crystal approximation}

	The thin crystal approximation assumes that the pump focal parameter $f_p$ is sufficiently small, given by the relation, $f_p = \frac{L}{2L_R} = \frac{L}{k_pw_p^2} \ll 1$, where $L$ is the crystal length, $L_R$ is the Rayleigh range, $k_p$ is the pump wave number, and $w_p$ is the pump beam waist. 
While this approximation is valid for thin crystals with millimeter-scale lengths \cite{Sala11}, it is unsuitable for systems requiring high brightness. 
Since brightness generally increases with the crystal length, centimeter-scale nonlinear crystals are typically used to generate bright photon sources \cite{Jab17}. 
Under such conditions, the thin crystal approximation breaks down.
Furthermore, under this approximation, it is not possible to define the focal parameters that maximize brightness. A previous study has suggested that brightness is maximized when the focal parameters of the signal and idler photons diverge to infinity \cite{Ling08}. 
However, this result contradicts the thin crystal approximation, indicating that the focal parameters maximizing brightness cannot be meaningfully defined under this condition. 
Thus, to analyze the focal parameters that optimize brightness in centimeter-scale crystals, a method beyond the thin crystal approximation is required.
     
     Given these limitations, we emphasize the need for a more general analytical framework for biphotons generated from type-0 pp crystals.
Specifically, our study addresses the conditions of non-degenerate states, broad spectral distributions, and long crystal lengths conditions that are critical for accurately characterizing SPDC processes in practical quantum photonic applications.

\section{\label{Appendix:B} Coincidence amplitude}
\renewcommand{\thefigure}{B\arabic{figure}}
\setcounter{figure}{0}

	In this appendix, we provide a detailed description of the method used to calculate and approximate the coincidence amplitude $C$.

\subsection{\label{Appendix:B-1}Calculation of coincidence amplitude}

        From Eq. (\ref{e5_6}), the coincidence amplitude $C$ can be written in terms of $\bm{k}$-space notation as follows:
            \begin{align}
                C_{n_s,n_i}^{l_s, l_i} &= \iint \bm{dq_s}\bm{dq_i}\, S_p(\omega_s+\omega_i) \widetilde{V_p}(\bm{q_s}+\bm{q_i})[\widetilde{LG}^{l_s}_{n_s} (\bm{q_s})]^*[\widetilde{LG}^{l_i}_{n_i} (\bm{q_i})]^*\int^{\frac{L}{2}}_{-\frac{L}{2}} dz\,\exp(i\Delta k_z z).
                \label{ec_1}
            \end{align}
        From the properties of the delta function $\delta (x)$, the following relations hold:
            \begin{align}
                \widetilde{V_p}(\bm{q_s}+\bm{q_i}) = \int \bm{dq_p} \, \delta(\bm{q_p}-\bm{q_s}-\bm{q_i})\widetilde{V_p}(\bm{q_p}),
                \label{ec_2}
            \end{align}
            \begin{align}
                \delta(q_{pa}-q_{sa}-q_{ia}) = \frac{1}{2\pi}\int da \exp(i(q_{pa}-q_{sa}-q_{ia})a), \, a \in \{x, y\}.
                \label{ec_3}
            \end{align}
        Using Eqs. (\ref{ec_1}), (\ref{ec_2}), and (\ref{ec_3}) along with the definition of the angular spectral amplitude \cite{Song20}, the coincidence amplitude $C$ is reformulated in terms of $\bm{x}$-space notation as follows:
            \begin{align}
                C_{n_s,n_i}^{l_s, l_i} = \frac{1}{4\pi^2}\int_0^{\infty} r\,dr \int_0^{2\pi} d\phi \, \int^{\frac{L}{2}}_{-\frac{L}{2}} dz \, S_p(\omega_s+\omega_i) V^{'}_p(r, z) [{LG^{'}}^{l_s}_{n_s}(r, z)]^*[{LG^{'}}^{l_i}_{n_i}(r, z)]^* \exp(i\Delta k z) \exp(i(l_p-l_s-l_i)\phi).
                \label{ec_4}
            \end{align}
        The integration over $r$ can be evaluated using the following equation \cite{Poh01}:
            \begin{align}
                &\int_0^{\infty} r^{\rho-1} \exp(-\sigma r) L_{n1}^{\alpha_1}(\lambda_1r)L_{n2}^{\alpha_2}(\lambda_2r) =  \binom{n_1+\alpha_1}{n_1} \binom{n_2+\alpha_2}{n_2}\frac{\Gamma(\rho)}{\sigma^{\rho}} F_A^{(2)}\left[\rho, -n_1, -n_2 ; \alpha_1 +1, \alpha_2 +1; \frac{\lambda_1}{\sigma}, \frac{\lambda_2}{\sigma}\right],
                \label{ec_5}
            \end{align}
        where $F_A$ is the first Lauricella hypergeometric function. 
        By applying the conditions in Sec.~\ref{sec:level2}, $C$ can further be expressed solely as an integral over $z$ ($= \frac{L}{2}u$):
            \begin{align}
                C^l_{n_s, n_i}=& \sum_{m_s=0}^{n_s}\sum_{m_i=0}^{n_i} \alpha_{m_s, m_i}^{l, n_s, n_i} \int_{-1}^1\,du\,G^{l, n_s, n_i}_{m_s, m_i}(u)\,\exp(i \Phi u),
                \label{ec_6}
            \end{align}
        The coefficients $\alpha_{m_s, m_i}^{l, n_s, n_i}$, $G^{l, n_s, n_i}_{m_s, m_i}$, and $g^{*}(g_p, g_s, g_i)$ are given by:
            \begin{align}
                \alpha_{m_s, m_i}^{l, n_s, n_i} &= \frac{L2^{l+m_s+m_i-\frac{3}{2}}(-1)^{m_s+m_i}}{\pi^{\frac{5}{2}}} \frac{\sqrt{n_s!n_i!(n_s+l)!(n_i+l)!}(l+m_s+m_i)!}{(n_s-m_s)!(l+m_s)!m_s!(n_i-m_i)!(l+m_i)!m_i!}\frac{w_p^{2m_s+2m_i+2l+1}w_s^{2m_i+l+1}w_i^{2m_s+l+1}}{(w_p^2w_s^2+w_p^2w_i^2+w_s^2w_i^2)^{m_s+m_i+l+1}},
		\label{ec_7}
            \end{align}

            \begin{align}
                G^{l, n_s, n_i}_{m_s, m_i} &= \frac{g_p^{m_s+m_i+l}g_s^{n_s-m_s}g_i^{n_i-m_i}}{{g_s^*}^{n_s-m_i}{g_i^*}^{n_i-m_s}{g^{*}(g_p, g_s, g_i)}^{m_s+m_i+l+1}} ,
                \label{ec_8}
            \end{align}
        and
            \begin{align}
		g^{*}(g_p, g_s, g_i) = \frac{w_p^2w_s^2g_pg_s^* + w_p^2w_i^2g_pg_i^* + w_s^2w_i^2g_s^*g_i^*}{w_p^2w_s^2+w_p^2w_i^2+w_s^2w_i^2}.
                \label{ec_9}
            \end{align}

\subsection{\label{Appendix:B-2} Approximation of coincidence amplitude}

        Unlike $w_p$, which has a one-to-one correspondence with $f_p$ under the monochromatic pump condition, the signal and idler waists $w_s$ and $w_i$ do not uniquely correspond to $f_s$ and $f_i$ due to frequency variations.

\subsubsection{\label{B-2-a}Approximation conditions}

            \begin{figure}[htbp]
            \centering
                \includegraphics[width=0.9\textwidth]{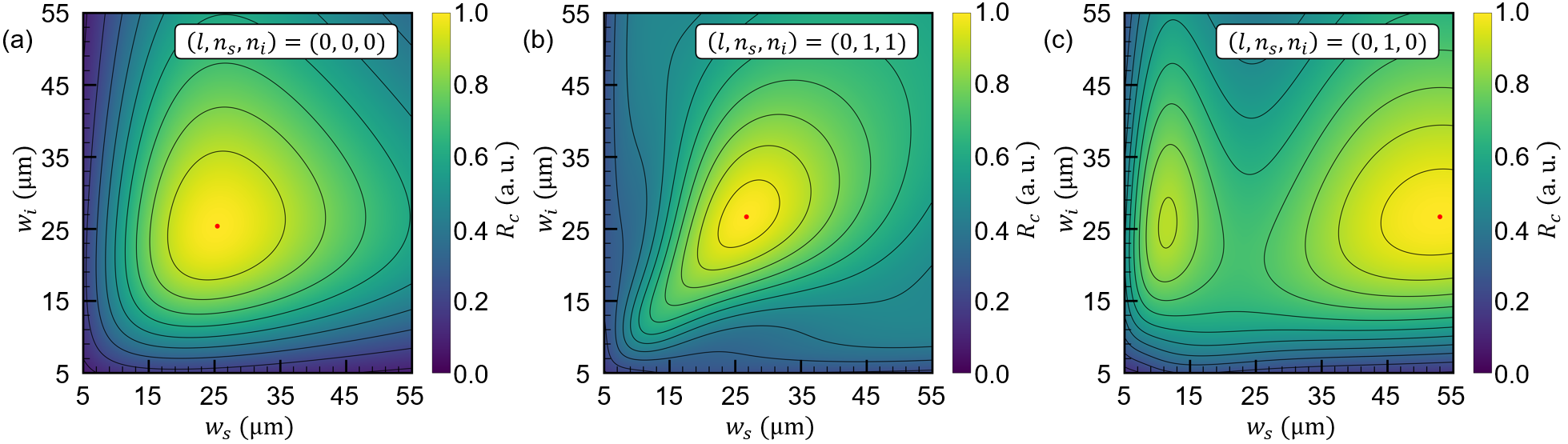}
                \caption{Normalized pair collection rate $R_c$ as a function of the signal waist $w_s$ and idler waist $w_i$ for different $(l,n_s,n_i)$ configurations. The red dots indicate the $(w_s^{opt}, w_i^{opt})$ at which $R_c$ reaches its maximum at $T = 24.5\,^{\circ}\mathrm{C}$. (a) $(l,n_s,n_i)=(0,0,0)$, $(w_s^{opt}, w_i^{opt})=(25.4\,\mathrm{\upmu m},25.4\,\mathrm{\upmu m})$. (b) $(l,n_s,n_i)=(0,1,1)$, $(w_s^{opt}, w_i^{opt})=(26.7\,\mathrm{\upmu m},26.7\,\mathrm{\upmu m})$. (a) $(l,n_s,n_i)=(0,1,0)$, $(w_s^{opt}, w_i^{opt})=(53.0\,\mathrm{\upmu m},26.7\,\mathrm{\upmu m})$.}
                \label{figb1}
            \end{figure}

	To simplify the coincidence amplitude $C$, we propose two conditions:

\begin{enumerate}
    \item
    Previous studies \cite{Miat11} indicate that under the condition $w_{s}=w_{i}$, the coincidence probability $P_{n_{si}}^{l}$ is maximized when $n_{s}=n_{i}$. 
    We confirm that the pair collection rate $R_c$ is also maximized under this condition.
    As shown in Figs. \ref{figb1} (a) and (b), Under the condition $n_s=n_i$, the maximum $R_c$ is achieved when $w_s=w_i$. 
    Therefore, we proceed with calculations under $n_s=n_i=n_{si}$ and $w_s=w_i=w_{si}$.
    If the radial indices of the signal and idler are not identical, the optimized waists of the signal and idler that maximize $R_c$ are not identical, as shown in Fig. \ref{figb1} (c).

    \item
    We assume that $k_s$ and $k_i$ deviate approximately symmetrically from the degenerate wave number $k_d=k(\frac{\omega_p}{2})$, which leads to the following relation:
                \begin{align}
                    (k_s, k_i) \approx (k_d + \Delta k, k_d - \Delta k).
                    \label{ec_10}
                \end{align}
    For type-0 SPDC, where signal and idler photons share polarization \cite{Stein14}, their refractive indices are nearly identical, validating this assumption. 
    Additionally, for small $\left(\frac{\Delta k} {k_d}\right)^2\ll1$, higher-order term $O\left(\left(\frac{\Delta k} {k_d}\right)^2\right)$ is negligible and can be omitted.
\end{enumerate}

To further simplify $g(g_p, g_s, g_i)$, we rewrite it as:
            \begin{align}
                g(g_p, g_s, g_i) &= (1+if_1u)(1+if_2u),
                \label{ec_26}
            \end{align}
	where $f_1, f_2$ satisfy:
            \begin{align}
                f_1 + f_2 &= \frac{L}{w_p^2w_s^2+w_p^2w_i^2+w_s^2w_i^2}\left(\frac{w_p^2+w_i^2}{k_s}+\frac{w_p^2+w_s^2}{k_i}-\frac{w_s^2+w_i^2}{k_p}\right), \notag \\ 
                f_1f_2 &= L^2\frac{(k_p-k_s-k_i)}{k_pk_sk_i}.
                \label{ec_28}
            \end{align}

\subsubsection{\label{B2b}Approximation of $f_1$ and replacement of $g_s$ and $g_i$}

	Without loss of generality, we assume $f_1>f_2$. 
According to a previous study \cite{Ben10}, the term $f_2$ can be neglected under the following conditions: crystal length $L\gtrsim$1 mm, refractive index $n\gtrsim$ 1.5, poling period $\Lambda \gtrsim$ 5 $\mathrm{\upmu m}$, signal or idler wave length $\lambda_{s(i)}\lesssim$ 1.6 $\mathrm{\upmu m}$, and pump wavelength $\lambda_p\lesssim$ 0.8 $\mathrm{\upmu m}$.
As our calculation conditions satisfy these criteria, we can neglect $f_2$ and approximate $f_1$ as
            \begin{align}
                f_1 &\approx \frac{L}{w_p^2w_s^2+w_p^2w_i^2+w_s^2w_i^2}\left(\frac{w_p^2+w_i^2}{k_s}+\frac{w_p^2+w_s^2}{k_i}-\frac{w_s^2+w_i^2}{k_p}\right).
                \label{ec_29}
            \end{align}
	
Considering the two conditions outlined in Appendix B. 2. \ref{B-2-a} and applying them to Eq. (\ref{ec_9}), $f_1$ can be further approximated as 
            \begin{align}
                f_1 &\approx \frac{2L}{k_d w_{si}^2} \frac{k_p(1+\gamma^2)-k_d\left(1-\left(\frac{\Delta k}{k_d}\right)^2\right)}{k_p(1+2\gamma^2)\left(1-\left(\frac{\Delta k}{k_d}\right)^2\right)} \approx \underbrace{\frac{L}{k_d w_{si}^2}}_{f^d_{si}} \underbrace{\frac{2(k_p(1+\gamma^2)-k_d)}{k_p(1+2\gamma^2)}}_{h(k_p, k_d, \gamma)},
                \label{ec_11}
            \end{align}
	where $\gamma = \frac{w_p}{w_{si}}$. 
As $\gamma$ increases from 0 to $\infty$, the function $h(k_p, k_d, \gamma)$ satisfies
            \begin{align}
                1 \le h(k_p, k_d, \gamma) < \frac{1+\frac{\delta k}{k_d}}{1+\frac{\delta k}{2k_d}},
                \label{ec_12}
            \end{align}
	where $\delta k = k_p - 2k_d$. 
Under our calculation conditions, $\frac{\delta k}{k_d} < 0.127$, implying $h(k_p, k_d, \gamma) < 1.06$. 
Therefore, we can approximate $f_1$ as $f^d_{si}$ with an error margin of less than 6\%, which leads to the approximation $g(g_p, g_s, g_i)\approx g^d_{si} = 1 + i f^d_{si} u$.

\subsubsection{\label{B2c}Approximation of $g_s$ and $g_i$}

Not only can $g(g_p, g_s, g_i)$ be approximated as $g^d_{si}$, but the individual terms $g_s$ and $g_i$ can also be replaced due to symmetry. 
Given the conditions established earlier, we redefine $g_s$ and $g_i$ as follows:
            \begin{align}
                g_s &= 1+i\frac{L}{k_sw_{si}^2}u \approx 1+i\frac{L}{k_d\left(1+\frac{\Delta k}{k_d}\right)w_{si}^2}u \approx 1+if^d_{si}\left(1-\frac{\Delta k}{k_d}\right)u, \notag \\ 
                g_i &\approx 1+if^d_{si}\left(1+\frac{\Delta k}{k_d}\right)u.
                \label{ec_13}
            \end{align}
	The product terms can then be expressed as follows:
            \begin{align}
                g_sg_i &\approx 1+if^d_{si}(2u) +(if^d_{si}u)^2\left(1-\left(\frac{\Delta k}{k_d}\right)^2\right) \approx (1+if^d_{si}u)^2 = {g^d_{si}}^2,\notag \\
                g_sg_s^* &\approx 1+\left(f^d_{si}\left(1-\frac{\Delta k}{k_d}\right)u\right)^2 \approx 1+{f^d_{si}}^2\left(1-2\frac{\Delta k}{k_d}\right)u^2, \notag \\
                g_ig_i^* &\approx 1+{f^d_{si}}^2\left(1+2\frac{\Delta k}{k_d}\right)u^2.
                \label{ec_14}
            \end{align}

Since $n_s=n_i=n_{si}$, for any $\alpha^{l, n_{si},n_{si}}_{m_s^{'}, m_i^{'}}G^{l, n_{si},n_{si}}_{m_s^{'}, m_i^{'}}$ corresponding to arbitrary $(m_s, m_i) = (m_s^{'}, m_i^{'})$ in Eq. (\ref{ec_6}), there always exists a corresponding term $\alpha^{l, n_{si},n_{si}}_{m_i^{'}, m_s^{'}}G^{l, n_{si},n_{si}}_{m_i^{'}, m_s^{'}}$ corresponding to $(m_s, m_i) = (m_i^{'}, m_s^{'})$.
Furthermore, from Eq. (\ref{ec_7}), it follows that $\alpha^{l, n_{si},n_{si}}_{m_s^{'}, m_i^{'}}=\alpha^{l, n_{si},n_{si}}_{m_i^{'}, m_s^{'}}$. 
Thus, Eq. (\ref{ec_6}) can be rewritten as
            \begin{align}
                (C^l_{n_{si}, n_{si}}=) C^l_{n_{si}}= \sum_{m_s=0}^{n_{si}}\sum_{m_i=0}^{n_{si}} \alpha_{m_s, m_i}^{l, n_{si}, n_{si}} \int_{-1}^1\,du\,\frac{G^{l, n_{si}, n_{si}}_{m_s, m_i}+G^{l, n_{si}, n_{si}}_{m_i, m_s}}{2}\,\exp(iPu).
                \label{ec_25}
            \end{align}

        Without loss of generality, let $m_s\,-\,m_i\,=\,m \geq 0$. 
Then, the sum of $G^{n_{si}, n_{si}}_{m_s, m_i}$ and $G^{n_{si}, n_{si}}_{m_i, m_s}$ is given by
            \begin{align}
                G^{l, n_{si}, n_{si}}_{m_s, m_i} + G^{l, n_{si}, n_{si}}_{m_i, m_s} &= \frac{g_p^{m_s+m_i+l}}{{g_d^*}^{m_s+m_i+l+1}}\underbrace{\left(\frac{g_s^{n_{si}-m_s}g_i^{n_{si}-m_i}}{{g_s^{*}}^{n_{si}-m_i}{g_i^*}^{n_{si}-m_s}}+\frac{g_s^{n_{si}-m_i}g_i^{n_{si}-m_s}}{{g_s^{*}}^{n_{si}-m_s}{g_i^*}^{n_{si}-m_i}}\right)}_{I(n_{si}, m_s, m_i)}.
                \label{ec_15}
            \end{align}
        Applying the approximation from Eq. (\ref{ec_14}), we obtain
            \begin{align}
                I(n_{si}, m_s, m_i) &= \frac{(g_sg_i)^{n_{si}-m_s}}{(g_s^*g_i^*)^{n_{si}-m_i}}\left(\frac{(g_ig_i^*)^m+(g_sg_s^*)^m}{(g_s^*g_i^*)^m}\right) \notag \\
                &\approx \frac{g_d^{2(n_{si}-m_s)}}{{g_d^*}^{2(n_{si}-m_i)}}\underbrace{\left(\left(1+{f^d_{si}}^2\left(1+2\frac{\Delta k}{k_d}\right)u^2\right)^m + \left(1+{f^d_{si}}^2\left(1-2\frac{\Delta k}{k_d}\right)u^2\right)^m\right)}_{J(f^d_{si}, \frac{\Delta k}{k_d}, u)}.
                \label{ec_16}
            \end{align}
        Defining $J(f^d_{si},\frac{\Delta k}{k_d}, u)$, we approximate it to eliminate its dependency on $\frac{\Delta k}{k_d}$:
            \begin{align}
                J(f^d_{si}, \frac{\Delta k}{k_d}, u) &= (1+{f^d_{si}}^2u^2)^m + m(1+{f^d_{si}}^2u^2)^{m-1}\left({f^d_{si}}\frac{2\Delta k}{k_d}u^2\right) + O\left(\left(\frac{\Delta k}{k_d}\right)^2\right), \notag \\
                &+ (1+{f^d_{si}}^2u^2)^m + m(1+{f^d_{si}}^2u^2)^{m-1}\left(-{f^d_{si}}\frac{2\Delta k}{k_d}u^2\right) + O\left(\left(\frac{\Delta k} {k_d}\right)^2\right), \notag \\
                &\approx 2(1+{f^d_{si}}^2u^2)^m = 2({g^d_{si}}{g^d_{si}}^*)^m.
                \label{ec_17}
            \end{align}
       As a result of Eq. (\ref{ec_16}) and Eq. (\ref{ec_17}), Eq. (\ref{ec_15}) simplifies to
            \begin{align}
                \frac{G^{l, n_{si}, n_{si}}_{m_s, m_i} + G^{l, n_{si}, n_{si}}_{m_i, m_s}}{2} \approx \frac{g_p^{m_s+m_i+l}{g^d_{si}}^{2n_{si}-m_s-m_i}}{{{g^d_{si}}^*}^{2n_{si}+l+1}}.
                \label{ec_18}
            \end{align}
        This result implies that in Eq. (\ref{ec_8}), the terms $g_s$, $g_i$, and $g(g_p, g_s, g_i)$ in $G^{l, n_{si}, n_{si}}_{m_s, m_i}$ can be effectively replaced with $g^d_{si}$. 
Furthermore, the coefficient $\alpha_{m_s, m_i}^{l, n_{si}, n_{si}}$ is given by
            \begin{align}
		\alpha_{m_s, m_i}^{l, n_{si}, n_{si}} = \frac{L2^{l+m_s+m_i-\frac{3}{2}}(-1)^{m_s+m_i}}{\pi^{\frac{5}{2}}}\frac{n_{si}!(n_{si}+l)!(l+m_s+m_i)!}{(n_{si}-m_s)!(l+m_s)!m_s!(n_{si}-m_i)!(l+m_i)!m_i!}\frac{1}{w_p\left(2+\frac{1}{\gamma^2}\right)^{m_s+m_i+l+1}}.
                \label{ec_21}
            \end{align}
	Next, approximating $\frac{1}{\gamma^2} = \frac{f_p}{f^d_{si}}\left(2+\frac{\delta k}{k_d}\right)$ as $\frac{2f_p}{f^d_{si}}$, which introduces an error of less than 6$\%$, and substitute the beam waist parameter with $w_p=\sqrt{\frac{L}{k_pf_p}}$.
Under these approximations, the coefficient $\alpha_{m_s, m_i}^{l, n_{si}, n_{si}}$ can be approximated as a new coefficient $\beta^{l, n_{si}}_{m_s, m_i}$, defined as
            \begin{align}
		\beta^{l, n_{si}}_{m_s, m_i} =  \frac{\sqrt{Lk_p}(-1)^{m_s+m_i}}{(2\pi)^{\frac{5}{2}}}\frac{\sqrt{f_p}}{\left(1+\frac{f_p}{f^d_{si}}\right)^{m_s+m_i+l+1}}\frac{n_{si}!(n_{si}+l)!(l+m_s+m_i)!}{(n_{si}-m_s)!(l+m_s)!m_s!(n_{si}-m_i)!(l+m_i)!m_i!}.
                \label{ec_27}
            \end{align}
Thus, Eq. (\ref{ec_6}) can be approximated as
            \begin{align}
                C^l_{n_{si}} \approx \sum_{m_s=0}^{n_{si}}\sum_{m_i=0}^{n_{si}} \beta^{l, n_{si}}_{m_s, m_i} \int_{-1}^1\,du\,{G^d}^{l, n_{si}}_{m_s, m_i}(u)\,\exp(i \Phi u),
                \label{ec_19}
            \end{align}
        where
            \begin{align}
                {G^d}^{l, n_{si}}_{m_s, m_i} = \frac{g_p^{m_s+m_i+l}{g^d_{si}}^{2n_{si}-m_s-m_i}}{{{g^d_{si}}^*}^{2n_{si}+l+1}}\textcolor{red}{.}
                \label{ec_20}
            \end{align}
From Eq. (\ref{ec_20}) and Eq. (\ref{ec_27}), it follows that both ${G^d}^{l, n_{si}}_{m_s, m_i}$ and $\beta^{l, n_{si}}_{m_s, m_i}$ are solely determined by the pump focal parameter $f_p$ and the degenerate focal parameter $f^d_{si}$.
Notably, they are independent of the signal-idler frequency $\omega_{si}$. 
Consequently, the integral over $\omega_{si}$ in the pair collection rate $R_c$, given by Eq. (\ref{ec_30}), can be expressed solely as an integration over the phase mismatch term $\Phi$, which is denoted as $Q(T, u_1-u_2)$.
            \begin{align}
		R_c &= \int d\omega_{si}\,P^l_{n_{si}} \notag \\ 
		&\approx \frac{Lk_p}{(2\pi)^5}\sum_{m_{s1}=0}^{n_{si}}\sum_{m_{i1}=0}^{n_{si}}\sum_{m_{s2}=0}^{n_{si}}\sum_{m_{i2}=0}^{n_{si}}\frac{f_p}{\left(1+\frac{f_p}{f^d_{si}}\right)^{m_{s1}+m_{i1}+m_{s2}+m_{i2}+2l+2}}\zeta^{l, n_{si}}_{m_{s1}, m_{i1}}\zeta^{l, n_{si}}_{m_{s2}, m_{i2}} \notag \\ 
		&\quad \times \int_{-1}^1\,du_{1}\,\int_{-1}^1\,du_{2}\, {G^d}^{l, n_{si}}_{m_{s1}, m_{i1}}(u_1) {{G^{*d}}^{l, n_{si}}_{m_{s1}, m_{i1}}}(u_2) \underbrace{\int d\omega_{si}\,\exp(i\Phi(u_1-u_2))}_{Q(T, u_1-u_2)},
		\label{ec_30}
            \end{align}
where the coefficient $\zeta$ is expressed as
            \begin{align}
		\zeta^{l, n_{si}}_{m_s, m_i} =  \frac{(-1)^{m_s+m_i}n_{si}!(n_{si}+l)!(l+m_s+m_i)!}{(n_{si}-m_s)!(l+m_s)!m_s!(n_{si}-m_i)!(l+m_i)!m_i!}.
		\label{ec_31}
            \end{align}
Since $Q(T, u_1-u_2)$ is independent of the focal parameters and depends solely on the temperature, it follows that both the coincidence amplitude $C^l_{n_{si}}$ and the pair collection rate $R_c$ are determined by the two focal parameters $f_p$ and $f^d_{si}$ at a fixed temperature.

\section{\label{Appendix:C} Temperature dependence of the conditional optimized degenerate focal parameter}
\renewcommand{\thefigure}{C\arabic{figure}}
\setcounter{figure}{0}

            \begin{figure}[ht!]
                \centering
                \includegraphics[width=0.8\textwidth]{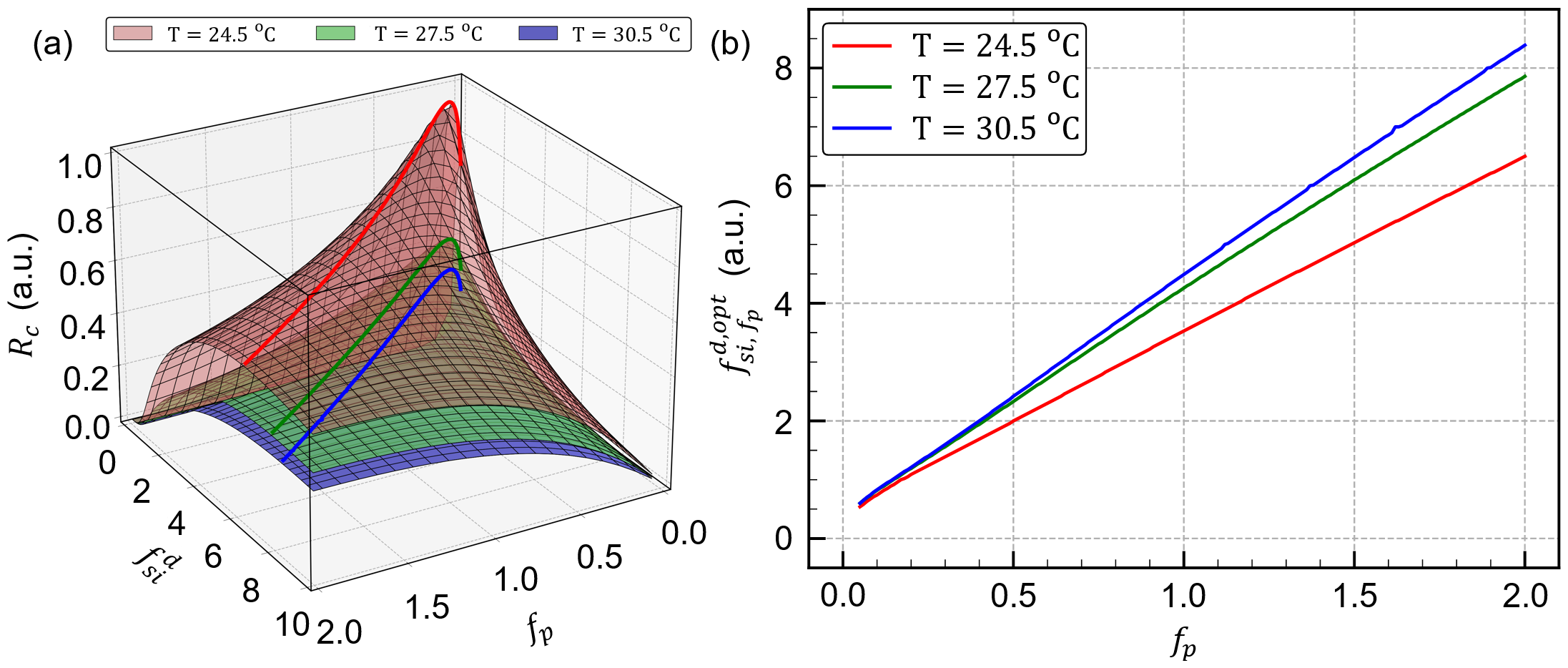}
                \caption{(a) Normalized pair collection rate $R_c$ as a function of the degenerate focal parameter $f^d_{si}$ and the pump focal parameter $f_p$ for three values of $T : 24.5\, ^{\circ}\mathrm{C}, 27.5\, ^{\circ}\mathrm{C} $, and $30.5\, ^{\circ}\mathrm{C}$ with $(l, n_{si}) = (2, 0)$. The solid lines indicate $R^{max}_{c, f_p}$ and $f^{d, opt}_{si, f_p}$. (b) $f^{d, opt}_{si, f_p}$ as a function of the pump focal parameter $f_p$.}
                \label{figc1}
            \end{figure}

 As shown in Fig.~\ref{figc1}(b), $f_{si, f_p}^{d, opt}$ for the same $f_p$ exhibits an increasing trend with temperature $T$. 
To analyze the effect of $T$ on $f_{si, f_p}^{d, opt}$, we reformulate the pair collection rate $R_c$ as
	\begin{align}
		R_c = \int d\Phi \, \frac{d\omega_{si}}{d\Phi} \, P_{n_{si}}^{l}.
		\label{ec_22}
	\end{align}

	We first examine how the distribution of $\frac{d\omega_{si}}{d\Phi}$ as a function of $\Phi$ evolves with increasing $T$.
	Using the quadratic approximation of the wavenumber $k$ as a function of frequency $\omega$ \cite{Bag22},we express $\frac{d\omega_{si}}{d\Phi}$ as
	\begin{align}
		\frac{d\omega_{si}}{d\Phi} = \pm \frac{1}{\sqrt{\frac{L^2}{4}\left(-\frac{1}{u_{g,s}}+\frac{1}{u_{g,i}}\right)+L(G_s+G_i)\Phi}},
		\label{ec_23}
	\end{align}
	where $u_g = 1/(\partial k/\partial \omega)$ is the group velocity, and $G = \partial/\partial\omega(1/u_g)$ represents the group velocity dispersion.
	The parameters $u_{g,s}$, $u_{g,i}$, $G_s$, and $G_i$ are evaluated at the signal and idler frequencies that satisfy the phase-matching condition $\Delta k = 0$. 
For high-order LG modes, $\Phi$ is generally non-positive ($\Phi \leq 0$) \cite{Hua18}. 
Therefore, it is sufficient to consider $\frac{d\omega_{si}}{d\Phi}$ only in the region where $\Phi \leq 0$.
From Eq. (\ref{ec_23}), it follows that the absolute value of $\frac{d\omega_{si}}{d\Phi}$ decreases as the absolute value of $\Phi$ increases.
To quantify this effect, we define $\Phi_{1/2}$ as the value of $\Phi$ at which $\frac{d\omega_{si}}{d\Phi}$ reduces to half of its value at $\Phi=0$, which can be expressed as
	\begin{align}
		\Phi_{\frac{1}{2}} = -\frac{L(-\frac{1}{u_{g,s}}+\frac{1}{u_{g,i}})}{16(G_s + G_i)}.
		\label{ec_24}
	\end{align}

	As the temperature increases, the difference between the signal and idler frequencies satisfying $\Delta k = 0$ becomes larger.
Consequently, the difference between the group velocities of the signal and idler, $u_{g,s}$ and $u_{g,i}$, also increases.
This leads to an increase in $\Phi_{1/2}$, as described by Eq. (\ref{ec_24}).
As a result, the distribution of $\frac{d\omega_{si}}{d\Phi}$ as a function of $\Phi$ shifts further from $\Phi=0$ with increasing $T$.
This indicates that to maximize $R_c$, the distribution of $|C^{l}_{n_{si}}|^2$ as a function of $\Phi$ must also shift further from $\Phi=0$.
This implies that the spectral distribution of the coincidence probability must shift further from the central frequency ($\omega_r = 1$).
As shown in Sec.~\ref{sec:level2}, such a shift in the spectral distribution can only be achieved by increasing $f_{si}^{d}$.
Therefore, we conclude that $f_{si, f_p}^{d, opt}$ for the same $f_p$ must increase as $T$ rises.

\twocolumngrid

\end{document}